\documentclass[12pt]{article}
\usepackage{graphicx}

\begin{document}
\title{Chiral expansion of the $\pi^0\rightarrow\gamma\gamma$ decay width }
\author{Bing An Li\\
Department of Physics and Astronomy, University of Kentucky\\
Lexington, KY 40506, USA}

\maketitle
\begin{abstract}
A chiral field theory of mesons has been applied to
study the contribution of the current quark masses to the $\pi^0\rightarrow\gamma\gamma$ decay width at the
next leading order. $2\%$ enhancement has been predicted and there is no new parameter. 
\end{abstract}
\newpage
It is well known that the decay amplitude of $\pi^0\rightarrow\gamma\gamma$
is an exact result of the Adler-Bell-Jackiw(ABJ)[1] triangle anomaly in the leading order of chiral expansion, 
$m^2_\pi\rightarrow 0$
\begin{equation}
\Gamma(\pi^0\rightarrow\gamma\gamma) = \frac{\alpha^2 m^3_\pi}{16\pi^3 f^2_\pi} = 7.72\; \textrm{eV},
\end{equation}
where the pion decay constant $f_\pi=185\textrm{MeV}$ is taken. 
Eq. (1) is also revealed from the Wess-Zumino-Witten(WZW) anomaly [2] in the leading order of chiral expansion.
The measurements of the $\Gamma(\pi^0\rightarrow\gamma\gamma)$ have a long history. The decay width of the 
$\pi^0\rightarrow\gamma\gamma$ is listed [3] as 
\begin{equation}
\Gamma(\pi^0\rightarrow\gamma\gamma) = 7.74 (1 \pm 0.059)\; \textrm{eV}.
\end{equation} 
Recently, PrimEx Collaboration has reported a new accurate measurement [4] of the decay width of $\pi^0\rightarrow\gamma\gamma$
\begin{equation}
\Gamma(\pi^0\rightarrow\gamma\gamma) = 7.82 \pm 0.14(stat.) \pm 0.17(syst.)\; \textrm{eV}.
\end{equation}
With the $2.8\%$ total uncertainty, this result is a factor of 2.5 more precise than Eq. (2) [4].

The ABJ anomaly is at the leading order in chiral expansion.
The study of the
effects at the next leading order in chiral expansion has attracted a lot of attention for a long time [5,6]. 
In Refs. [5] 
within the framework of the Chiral Perturbation Theory (ChPT) by calculating the loop diagrams 
of the WZW anomaly [2] an enhancement to $\Gamma(\pi^0\rightarrow\gamma\gamma)$ has 
been found. In Ref. [6] the difference $f_{\pi^0}-f_{\pi^+}$ and the $\pi^0-\eta$ mixing are taken into account in 
calculating $\Gamma(\pi^0\rightarrow\gamma\gamma)$. 

In this paper the decay amplitude of the $\pi^0\rightarrow\gamma\gamma$ is computed to the next leading order in chiral expansion 
by using a chiral field theory of pseudoscalar, vector, and axial-vector mesons [7].   
In the case of two flavors the Lagrangian of the theory [7] is expressed as
\begin{eqnarray}
{\cal L}=\bar{\psi}(x)(i\gamma\cdot\partial+\gamma\cdot v
+\gamma\cdot a\gamma_{5}+eQ\gamma\cdot A
-mu(x))\psi(x)-\bar{\psi(x)}M\psi(x)\nonumber \\
-{1\over4}F^{\mu\nu}F_{\mu\nu}
+{1\over 2}m^{2}_{0}(\rho^{\mu}_{i}\rho_{\mu i}+
\omega^{\mu}\omega_{\mu}
+a^{\mu}_{i}a_{\mu i}
+f^{\mu}f_{\mu})
\end{eqnarray}
where M is the current quark mass matrix
\[\left(\begin{array}{c}
         m_{u}\hspace{0.5cm}0\\
         0\hspace{0.5cm}m_{d}
        \end{array}  \right ),\]
\(v_{\mu}=\tau_{i}\rho^{i}_{\mu} + \omega_{\mu}\),
\(a_{\mu}=\tau_{i}a^{i}_{\mu}+f_{\mu}\),
and \(u=exp\{i\gamma_{5}(\tau_{i}\pi_{i}+ \eta)\}\).
The parameter m is originated in the quark condensate [7]. 
The Lagrangian (4) is explicit chiral symmetric in the limit, 
$m_q\rightarrow 0$. Dynamical chiral symmetry breaking is embed
in the Lagrangian (4).

By integrating out the quark fields the Lagrangian of mesons is
obtained [7]. This procedure is equivalent to the quark loop calculation.
The kinetic terms of the mesons are generated by the quark loops. There are real and imaginary two parts.
The real part is used to normalize the fields of Eq. (4) to physical meson fields and describes the meson processes 
with normal parity and the imaginary part 
describes the anomaly. The Lagrangian of the ChPT [8] is revealed from the real part of the Lagrangian at low energies
and all the 10 coefficients of the ChPT are predicted [9].
At the $O(p^4)$ the imaginary part of the Lagrangian is derived [10] in which the $\omega$ and $\rho$ fields are involved 
\begin{eqnarray}
\lefteqn{{\cal L}_{\omega}=
\frac{N_{c}}{(4\pi)^{2}}{2\over 3}\varepsilon^{\mu\nu\alpha\beta}
\omega_{\mu}Tr\partial_{\nu}UU^{\dag}\partial_{\alpha}UU^{\dag}
\partial_{\beta}UU^{\dag}}\nonumber \\
 & &+\frac{2N_{c}}{(4\pi)^{2}}\varepsilon^{\mu\nu\alpha\beta}
\partial_{\mu}\omega_{\nu}Tr\{i[\partial_{\beta}UU^{\dag}
(\rho_{\alpha}+a_{\alpha})-\partial_{\beta}U^{\dag}U(\rho_{\alpha}
-a_{\alpha})]\nonumber \\
& &-(\rho_{\alpha}+a_{\alpha})U(\rho_{\beta}-a_{\beta})
U^{\dag}-2\rho_{\alpha}a_{\beta}\}.
\end{eqnarray}
${\cal L}_{\omega}$ (5) is exact the same as the one 
presented by Kaymakcalan, Rajeev, and Schechter(KRS)[11] and Eq. (5) is part of the WZW Lagrangian. 

The $\pi$ field of Eq. (4) is normalized to the physical pion 
\begin{eqnarray}
\pi^i\rightarrow {2\over f_\pi}\pi^i.
\end{eqnarray}
The pion decay constant $f_\pi$ is proportional to m which is determined as [7]
\begin{eqnarray} 
m^2 = \frac{f^2_\pi}{6g^2}(1-{2c\over g})^{-1},\;\;
c = \frac{f^2_\pi}{2gm^2_\rho},
\end{eqnarray} 
g is a universal coupling constant of this theory, which is via the normalization of the vector fields introduced.
Using the decay rate of $\rho\rightarrow ee^+$,
it is determined to be 0.395 [7,12].
$f_{\pi}$, g, and the current quark masses are the inputs of this theory. 
In the leading order in the chiral expansion the cancellation between the two vertices of Eq. (4)
\begin{eqnarray}
{\cal L}_1 = -\frac{2im}{f_\pi}\bar{\psi}\tau_i\gamma_5\psi\pi^i,\\ 
{\cal L}_2 = {1\over2}m\bar{\psi}\psi\pi^2.
\end{eqnarray}
leads to 
\begin{equation}
m^2_{\pi}=-{4\over f^2_\pi}\langle 0|\bar{\psi}\psi |0 \rangle (m_u + m_d).
\end{equation}
Therefore, the pion is a Goldstone boson. 
The Vector Meson Dominance (VMD) is a natural result and many physical processes of mesons have been studied.
Theory agrees with the data well [7,12].
This theory is phenomenologically successful [7,12].

In Ref. [7] in the leading order of chiral expansion, $m_q\rightarrow 0$, 
the ABJ anomaly of 
the $\pi^0\rightarrow\gamma\gamma$ decay 
\[{\cal L}_{\pi^0\rightarrow \gamma\gamma} = -\frac{\alpha}{\pi f_\pi}\epsilon^{\mu\nu\alpha\beta}
\partial_\mu A_\nu\partial_\alpha A_\beta\]
is via the VMD 
\[\rho\rightarrow{1\over2}eg A,\;\;\;  
\omega\rightarrow{1\over6}eg A\] 
derived from the vertex $\pi\omega\rho$, which is revealed from Eq. (5),
\[{\cal L}_{\omega\rho\pi} = -\frac{N_C}{\pi^2 g^2 f_\pi}\epsilon^{\mu\nu\alpha\beta}\partial_\mu\omega_\nu\rho^0_\alpha
\partial_\beta\pi^0.\]

As mentioned above, the vertex ${\cal L}_{\omega\rho\pi}$ is derived from the KRS form of the WZW anomaly.
On the other hand, using the vertex (8) and the vertices
\[{1\over g}\bar{\psi}\tau^3\gamma_\mu\psi \rho^0_\mu,\;\;{1\over g}\bar{\psi}\gamma_\mu\psi\omega_\mu,\]
it can be derived from calculating the quark triangle loop diagram too.
In this paper using the L (4) directly, the triangle quark loops for the process $\pi^0\rightarrow\gamma\gamma$ are calculated to the 
next leading order in chiral expansion.
The theoretical approach exploited in this paper is consistent with the approach of Ref. [5].   
Besides the vertex (8) 
there is another one obtained from the mixing between the $a_1$ field and $\partial\pi$ [7] 
\begin{equation}
{\cal L} = - {c\over g}{2\over f_\pi}\bar{\psi}\tau_i\gamma_\mu\psi \partial^{\mu}\pi^i,
\end{equation}
which contributes to the triangle diagrams of the $\pi^0\rightarrow\gamma\gamma$ decay too. 

To the next leading order in current quark mass the amplitude of the $\pi^0\rightarrow\gamma\gamma$ determined 
by the vertex (8) is expressed as
\begin{equation}
T^{(1)} = \frac{2N_C}{3\pi}\frac{\alpha}{f_\pi}\epsilon_{\lambda\sigma\mu\nu}q^{\lambda}p^{\sigma}
\epsilon^{\mu}(1)\epsilon^{\nu}(2)
\{1+{1\over3m}(m_d-4m_u) + {1\over12}\frac{m^2_\pi}{m^2}\}.
\end{equation}
The amplitude determined by the vertex (11) is expressed as
\begin{equation}
T^{(2)} = - \frac{N_C}{9\pi}{c\over g}\frac{\alpha}{f_\pi}
\epsilon_{\lambda\sigma\mu\nu}q^{\lambda}p^{\sigma}\epsilon^{\mu}(1)\epsilon^{\nu}(2) 
\frac{m^2_\pi}{m^2}.
\end{equation}
The amplitude $T^{(2)}$ shows that the contribution of the vertex (11) is at the $O(m_q)$. 
At the leading order in chiral expansion only
the vertex (8) contributes to the $\pi^0\rightarrow\gamma\gamma$ decay and the $T^{(1)}$ is the amplitude 
predicted by the ABJ anomaly.
In Eqs. (12,13) the factor $N_C$ is obtained from the trace of the quark loop. Using the expression of $m^2$ (7),
the total amplitude is found to be
\begin{equation}
T = \frac{2N_C}{3\pi}\frac{\alpha}{f_\pi}\epsilon_{\lambda\sigma\mu\nu}q^{\lambda}p^{\sigma}\epsilon^{\mu}(1)\epsilon^{\nu}(2)
\{1 + {1\over3m}(m_d-4m_u) +\frac{m^2_\pi}{2f^2_\pi}g^2(1-{2c\over g})^2\}.
\end{equation}
In Eqs. (12-14) the $m^2_\pi$ is expressed by Eq. (10) and it is at the $O(m_q)$. In the chiral limit, 
$m_q\rightarrow 0$, the amplitude 
determined by the ABJ anomaly is recovered.
The decay width is obtained 
\begin{equation}
\Gamma(\pi^0\rightarrow\gamma\gamma) = \frac{\alpha^2}{16\pi^3}\frac{m^3_\pi}{f^2_\pi}\{1 + {1\over3m}(m_d-4m_u) 
+ \frac{m^2_\pi}{2f^2_\pi}g^2(1-{2c\over g})^2\}^2.
\end{equation}
The values of $m_d=5.05^{+0.75}_{-0.95}\; \textrm{MeV}$ and $m_u=2.49^{+0.75}_{-0.79}\; \textrm{MeV}$ are taken from Ref. [3].
$m=0.24\; \textrm{GeV}$ is determined by Eq. (7). 
\begin{equation}
{1\over3m}(m_d-4m_u)\sim -0.68\times10^{-2}
\end{equation}
is estimated and 
\begin{equation}
\frac{m^2_\pi}{2f^2_\pi}g^2(1-{2c\over g})^2 = 0.0167.
\end{equation}
Using the values of the corrections (16,17), it is obtained
\begin{equation}
\Gamma(\pi^0\rightarrow\gamma\gamma) = \frac{\alpha^2}{16\pi^3}\frac{m^3_\pi}{f^2_\pi} \times 1.02
= 7.87\; \textrm{eV}.
\end{equation}

The terms at the $O(m_q)$ in Eq. (14) leads to $2\%$ enhancement of the decay width. 
The theoretical result obtained by this approach agrees with the experimental data reported by PrimEX [4] 
within the experimental errors.

\end{document}